\documentclass[12pt,preprint]{aastex}

\def\arcsecpoint{$''\!.$}
\def\deg{$^{\circ}$}

\shorttitle{Physical Conditons in the NLR in Mrk 573}
\shortauthors{Kraemer et al.}

\begin{document}

\title{Physical Conditions in the Inner Narrow Line Region of the Seyfert 2 
Galaxy Markarian 573\altaffilmark{1}} 

\author{S. B. Kraemer\altaffilmark{2}, M.L. Trippe\altaffilmark{3},
D. M. Crenshaw\altaffilmark{3}, M. Mel\'endez\altaffilmark{4}, 
H.R. Schmitt\altaffilmark{5},
\& T.C. Fischer\altaffilmark{3}}
\altaffiltext{1}{Based on observations made with the NASA/ESA Hubble Space
Telescope, obtained from the Data Archive at the Space Telescope Science
Institute, which is operated by the Association of Universities for
Research in Astronomy, Inc., under NASA contract NAS 5-26555.}
\altaffiltext{2}{Institute for Astrophysics and Computational Sciences,
Department of Physics, The Catholic University of America, Washington, DC 20064; and
Code 667, Astrophysics Science Division, NASA Goddard Space Flight Center, Greenbelt,
MD 20771.} 
\altaffiltext{3}{Department of Physics and Astronomy, Georgia State University,
Astronomy Offices, Atlanta, GA 30303.}
\altaffiltext{4}{NASA Postdoctoral 
Program Fellow, Goddard Space Flight Center, Greenbelt, MD, 20771.}
\altaffiltext{5}{Remote Sensing Division, Naval Research Laboratory,
Washington, DC 20375; and Interferometrics, Inc., Herndon, VA 20171.}

\begin{abstract}
We have examined the physical conditions within a bright emission-line knot in the
inner narrow-line region (NLR) of the Seyfert 2 galaxy Mrk 573 using optical
spectra and photoionization models. The spectra were obtained with the {\it Hubble
Space Telescope}/Space Telescope Imaging Spectrograph, through the 0\arcsecpoint2
$\times$ 52\arcsecpoint0 slit, at a position angle of $-71.2$\deg~, 
with the G430L and G750M gratings. Comparing the spatial
emission-line profiles, we found [Fe~X] $\lambda$ 6734  barely
resolved, [O~III] $\lambda$5007 centrally peaked, but broader than
[Fe~X], and [O~II] $\lambda$3727 the most extended. Spectra of the
central knot were extracted from a region 1\arcsecpoint1 in extent,
corresponding to the full-width at zero intensity in the cross-dispersion
direction, of the knot. The spectra
reveal that [Fe~X] is broader in velocity width and blue-shifted compared with lines 
from less ionized species. Our estimate of the bolometric
luminosity indicates that the active galactic nucleus (AGN) is radiating at or above
its Eddington Luminosity, which is consistent with its identification
as a hidden Narrow-Line Seyfert 1. We were able to successfully match
the observed emission line ratios with a three-component 
photoionization model. Two components, one to account for the [O~III]
emission and another in which the [Fe~X] arises, are directly ionized by the AGN,
while [O~II] forms in a third component, which is ionized by a heavily absorbed continuum.
Based on our assumed ionizing continuum and the
model parameters, we determined that the two directly-ionized components
are $\sim$ 55 pc from the AGN. We have found similar radial distances for the
central knots in the Seyfert 2 galaxies Mrk 3 and NGC 1068, but much smaller
radial distances for the inner NLR in the Seyfert 1 galaxies NGC 4151 and NGC 5548. Although
in general agreement with the unified model, these results suggest
that the obscuring material in Seyfert galaxies extends out to at least tens of parsecs from the AGN.

\end{abstract}

\section{Introduction}

Active Galactic Nuclei (AGN) are thought to be powered by accretion of matter onto
super-massive black holes, which reside at the gravitational centers of the host galaxies.
Seyfert galaxies, relatively low luminosity (L$_{bol}$ $\lesssim$ several $\times$ 10$^{45}$ ergs s$^{-1}$), nearby
($z \lesssim 0.1$) AGN, are typically 
grouped into two classes (Khachikian \& Weedman 1974). The spectra of Seyfert 1 galaxies are characterized by
broad (full width half maximum, FHWM, $\gtrsim$ 1000 km s$^{-1}$) permitted lines, narrower (FWHM $\lesssim$
1000 km s$^{-1}$) forbidden lines, and strong, non-stellar optical and UV continua, while
the spectra of Seyfert 2's show narrow permitted and forbidden lines and optical and UV 
continua dominated by the host galaxy. Spectral polarimetry of Seyfert 2s (e.g. Antonucci \&
Miller 1985) revealed the presence of broad permitted lines and non-stellar continua
in polarized light. This discovery led to the  unified model for Seyfert
galaxies (Antonucci 1993), which posits that the difference between the two 
results from the orientation of the active region with respect
to the observer's line-of-sight. The broad emission line gas and continuum source is 
surrounded by a large column of dusty gas, which is along our line-of-sight to Seyfert 2s
and hence obscures our view of the central active region in those galaxies. 
Based on variability, the broad line region (BLR) ranges in size from several to tens of light days (e.g.
Peterson et al. 2004), while the region in which the forbidden lines and narrower components
of the permitted lines form, the narrow line region (NLR) can extend out to $\sim$ 1 kpc (e.g.
Pogge 1989). 

It is now generally believed that the NLRs of Seyfert galaxies are comprised of
gas that has been photoionized by the non-stellar continuum emitted by the
AGN (in this paper, we refer to the central source of ionizing radiation as the ``AGN''). The morphology of the NLR is often roughly biconical, with the apex
of the cones at the AGN, which suggests that the ionizing radiation is
collimated by absorbing material, possibly the putative torus or a disk-wind,
at small radial distances. However, the source of the NLR gas is not
well-established. In the case of the extended NLR, at radial distances of $\gtrsim$
1 kpc, the gas probably lies in the plane of the host galaxy
(e.g. Pogge 1989). If, for example, the angle between the axis of the black hole/accretion disk 
and that of the host galaxy is small, the ionizing radiation may not intersect
the disk of the host galaxy, and the NLR may be compact (see Schmitt \& Kinney
1996). The gas within the bicone in the inner few 100 pcs of the
NLR typically appears to be outflowing radially (e.g. Ruiz et al. 2005),  perhaps by radiation 
pressure (Das, Crenshaw, \& Kraemer 2007; Everett \& Murray 2007).

We previously found that slitless spectra of 
Seyfert galaxies,
obtained with {\it Hubble Space Telescope (HST)}/Space Telescope Imaging Spectrograph (STIS), often revealed the presence of bright knots
of [O~III] $\lambda$5007 emission in the inner tens of parsecs of their NLRs (Crenshaw \& Kraemer 2005).
Based on photoionization models, we (Crenshaw \& Kraemer 2005) suggested that the physical conditions,
i.e., ionization state and column density,
within the emission-line knot and the outflowing UV
absorbers (Kraemer et al. 2006) were similar in NGC 4151. Furthermore, the emission-line gas in the central knot in NGC 4151
must have a high covering factor (40 -- 50\%) to account for the emission-line luminosities, which is
consistent with estimates of the global covering factors of the UV absorbers determined from the
fraction of Seyfert 1s that possess
intrinsic UV absorption (e.g. Crenshaw et al. 1999; Dunn et al. 2007). Hence, it is possible that
the UV absorbers and the emission-line knots arise in the same component of gas, at least in the case
of NGC 4151 and, possibly, in other Seyfert galaxies.

One of the Seyferts possessing a bright central [OIII] knot is the Seyfert 2 galaxy
Mrk 573 (cz $=$ 5160
($\pm$10) km s$^{-1}$ from H~I 21-cm observations [Springob et al. 2005]; this corresponds to a distance $\approx$ 72.6 Mpc, 
assuming $H_{\rm o} = 71$ km s$^{-1}$ Mpc$^{-1}$). The host galaxy is classified as (R)SAB(rs). It possesses
a triple radio source (Ulvestad \& Wilson 1984), comprised of a central component and two lobes, lying
along a position angle of $\sim$ 125$^{\rm o}$. High excitation optical emission-line gas 
extends $\sim$ 4.8 kpc along the radio axis (Tsvetanov \& Walsh 1992), and Pogge \& de
Robertis (1995) found that the extended [O~III] and [N~II] $\lambda\lambda$ 6548, 6583
$+$ H$\alpha$ emission possessed a biconical morphology, similar to two well-studied Seyfert 2 galaxies,
NGC 1068 and Mrk 3. Interestingly, while spectropolarimetry of both NGC 1068 (Antonucci \& Miller 1985) and Mrk 3 (Schmidt \& Miller
1985) revealed broad permitted lines and non-stellar continuum emission, there was some
question as to whether Mrk 573 possessed a hidden BLR (Tran 2001; 2003; but, see also Kay 1994).
However, spectropolarimetric observations with {\it Subaru} (Nagao et al. 2004) definitively showed
the presence of broad H$\alpha$ (FWHM $\approx$ 3000 km s$^{-1}$), H$\beta$, and the Fe~II multiplet, confirming the presence
of a BLR in Mrk 573.

Strong high-ionization forbidden lines, such as [Fe~VII] $\lambda$6087, [Fe~X] $\lambda$6374, and [Ne~V] $\lambda\lambda$ 3346,3426, have been
detected in ground-based optical spectra of Mrk 573 (Storchi-Bergmann et al. 1996; Erkens, Appenzeller, \&
Wagner 1997), indicating that the NLR is more highly ionized than many Seyfert 2 galaxies (see Koski 1978).   
In fact, Storchi-Bergmann et al. found high-ionization gas $\sim$ 2.5 Kpc from the AGN, which requires that
the AGN is quite luminous, in spite of suggestions to the contrary (Nagao et al. 2004).
Mullaney \& Ward (2007) determined that [Fe~VII] $\lambda$6087, [Fe~X] $\lambda$6374, and [Fe~XI] $\lambda$7892 were blue-shifted with
respect to [O~III] and the Balmer lines, indicating that the high-ionization lines formed in a distinct
component of NLR gas. 

We have analyzed archival {\it HST}/STIS longslit spectra of Mrk 573; at a distance of 73.6 Mpc,  the 0\arcsecpoint05
STIS CCD pixel size covers $\sim$ 18 pc. In this paper, we present 
an analysis of physical conditions in the central emission-line knot, similar to that
done for the central ``hot-spot'' in NGC 1068 (Kraemer \& Crenshaw 2000a) and NGC 4151 (Crenshaw \& Kraemer 2007). 
The structure of the paper is as follows. In Section 2, we present the details of the
observations, including the line profiles, measurement of the line fluxes and correction
for extinction. In Section 3, we estimate the luminosity of the AGN. In Section 4, we present
the input parameters and results of our photoionization models. In Section 5, we discuss
the implications of these results, and our previous studies of the NLR, on our understanding
of the obscuration of the inner NLR in Seyferts. Finally, in Section 6, we summarize the results.

\section{Observations and Observational Results}

We retrieved STIS long-slit spectra and Wide Field and Planetary Camera 2 (WFPC2) images of Mrk~573 from the
Multimission Archive at the Space Telescope Science Institute (MAST). The
spectra were obtained on 2001 October 17 over two {\it HST} orbits with the
G430L and G750M gratings, at significantly different spectral resolutions
of 5.5 \AA\ and 1.1 \AA\ respectively. The 52 $\times$ 0\arcsecpoint2 slit
was used at a position angle of $-$71.2\deg\ for all of the STIS
observations, and was centered on the optical continuum peak before the
first spectrum was obtained. For the G750M spectra, three separate spectral
images were obtained with the optical continuum peak displaced
0\arcsecpoint25 along the slit on either side of the initial pointing. For
the the G430M spectra, two separate images were obtained, with one
displaced 0\arcsecpoint25 along the slit. We give additional details of
each observation, listed in chronological order, in Table 1.

We processed the STIS spectra with the IDL software developed at NASA's
Goddard Space Flight Center for the STIS Instrument Definition Team. Prior
to processing, we shifted the raw images in the cross-dispersion direction
to align the two-dimensional spectra and facilitate the removal of cosmic
ray hits. The zero-point of the wavelength scales were corrected using
wavelength-calibration exposures obtained after each observation. The
spectra for each grating were combined, geometrically rectified, and
flux-calibrated to produce a constant wavelength along each column and
fluxes in units of ergs s$^{-1}$ cm$^{-2}$ \AA$^{-1}$ per cross-dispersion
pixel.

We show the placement of the STIS slit on a WFPC2 [O~III] image of Mrk~573
in Figure 1. We determined the location of the optical continuum peak,
identified by a ``+'' in the figure, by aligning the WFPC2 [O~III] and
continuum images obtained on the same date (Schmitt et al. 2003). The slit
runs through a bright, spatially resolved, central knot of [O~III]
emission, similar to those seen in most nearby Seyfert galaxies (Crenshaw
\& Kraemer 2005), as well as several arc-like structures in the NLR that
are bright in [O~III].

Figure 2 shows portions of the calibrated spectral images. The optical
continuum peak is the dark horizontal band in these images, and it appears
to be spatially resolved. Most of the strong emission lines show extended
structure along the slit, with bright spots corresponding to the bright
knots in Figure 1. The emission lines show some interesting kinematics,
which we will address in a future paper (T. Fischer et al., in preparation).

In Figure 3, we show brightness profiles of two lines, [O~III]
$\lambda$5007 and [Fe~X] $\lambda$6374, and the continuum just redward of
the [O~III] emission. These were obtained by summing the columns that
encompass the entire feature, and plotting the fluxes in the
cross-dispersion direction, after subtracting the continuum profile from
the lines and normalizing to the peak flux. The continuum profile is much
broader than the other two and is dominated by light from the host galaxy,
at least at distances $>$ 0\arcsecpoint5 from the peak. Scattered radiation
from the hidden nucleus may contribute to the peak continuum emission, but
there is no clear evidence for this (such as broad Balmer emission). The
central [O~III] emission is significantly broader (FWHM $=$
0\arcsecpoint28) than a resolution element (FWHM $=$ 0\arcsecpoint1).
The bumps in the [O~III] profile
correspond to locations where the slit intercepts the arcs of emission
shown in Figure 1. The central [Fe~X] profile is just barely resolved (FWHM
$=$ 0\arcsecpoint17) and it shows no strong evidence for additional extended
emission, indicating that this high-ionization emission is much more
concentrated toward the nucleus. In Figure 4, we show the brightness
profile of [O~II] $\lambda$3727, which is even more spread out than that of
[O~III].

We extracted spectra of the central knot of emission from the spectral
images by combining the central 22 rows (1\arcsecpoint1), corresponding to
the full-width at zero intensity of the central knot. We show the resulting
G430L and G750M spectra in Figures 5 and 6. Lines from a wide range in
ionization state, from [O~I] to [Fe~X], are present, similar to those seen
in the central NLR knot of the Seyfert 2 galaxy NGC~1068 (Kraemer \&
Crenshaw 2000a).

We measured the fluxes of most of the emission lines by integrating over a
local baseline continuum. For blended lines, such as H$\alpha$ and
[N~II]$\lambda\lambda$6548, 6584, we used the [O~III] $\lambda$5007 line as
a template to deblend the lines. We then determined the reddening of the
emission lines from the observed H$\alpha$/H$\beta$ ratio, the Galactic
reddening curve of Savage \& Mathis (1979), and an intrinsic ratio of 2.9.
The reddening is E(B-V) $=$ 0.18 $\pm$ 0.03 mag; only 0.02 mag can be
attributed to our Galaxy (Schlegel, Finkbeiner, \& Davis 1998). We determined
uncertainties in the observed lines ratios from the sum in quadrature of
the errors from photon noise and different reasonable continuum placements.
We determined uncertainties in the dereddened ratios by propagating the
above errors along with those for the reddening. Table 2 gives the observed
and dereddened narrow-line ratios, relative to H$\beta$.

In Figure 7, we plot the radial velocity centroids of the emission lines
with respect to the heliocentric systemic velocity (cz $=$ 5160),
as a function of ionization potential. The ionization potentials and velocities
relative to systemic are given in Table 2. All of the lines are slightly
redshifted with respect to the systemic velocity, although the [Fe~X] line
shows the lowest redshift. We can determine the intrinsic velocity widths
of the lines at $\lambda$ $\geq$ 6300 \AA, because these lines are resolved
in the G750M grating. All of the these lines have FWHM $=$ 200 km s$^{-1}$
to within 20 s$^{-1}$, except for [Fe~X] $\lambda$6374, which has FWHM $=$
380 km s$^{-1}$. The velocity offset and larger FWHM of the [Fe~X]
line suggest that the high-ionization gas comes from a distinct component.

\section{Luminosity and Mass of the AGN}

Guainazzi, Matt, \& Perola (2005) determined that the X-ray source in Mrk 573 is 
obscured by ``Compton-thick'' matter ($N_{H}$ $>$ 1.6 $\times$ 10$^{24}$ cm$^{-2}$),
which is consistent with its non-detection by the {\it Swift}/Burst Alert Telescope
(Tueller et al. 2008). Therefore, the intrinsic X-ray luminosity can only be determined
via indirect means. As described by Melendez et al. (2008a), the [O~IV] 25.89$\mu$m line is
a reliable isotropic quantity. From the {\it Spitzer}/IRS spectrum of Mrk 573 (see Mel\'endez et
al. 2008b), the [O~IV] luminosity is $L_{[OIV]}$ $\approx$ 10$^{41.7}$ ergs cm$^{-2}$ s$^{-1}$. Using the linear regression fit 
to the [O~IV] and the 2--10 keV ($L_{2-10 keV}$) luminosities for Seyfert 1 galaxies calculated by
Mel\'endez et al. (2008a), which has an uncertainty of $\sim$ 0.5 dex, we estimate that $L_{2-10 keV}$ is $\sim$ 10$^{44.1 \pm 0.5}$ ergs cm$^{-2}$ s$^{-1}$.

Using on the spectral fitting by Padovani \& Rafanelli (1988), Awaki et al. (2001) calculated a bolometric correction
of 27.2 for L$_{2-10 keV}$, which is in general agreement with a more recent analysis by Vasudevan \& Fabian (2007).
From Padovani \& Rafanelli (1988), the standard deviation in the correction for Seyfert 1s is $\sim$ 0.39 dex. 
Based on our estimate of the L$_{2-10 keV}$, we obtain
$L_{bol}$ $\sim$ 10$^{45.5 \pm 0.6}$ ergs cm$^{-2}$ s$^{-1}$, with the uncertainty equal to that of the bolometric
correction and L$_{2-10 keV}$ added in quadrature. Using the stellar velocity dispersion
measured by Nelson \& Whittle (1995), Bian \& Gu (2006) derived a black hole mass for Mrk 573 of log~(M$_{BH}$/M$_\odot$)
$=$ 7.28, which yields an Eddington luminosity $=$ 10$^{45.4}$ ergs cm$^{-2}$ s$^{-1}$. Therefore, the source appears
to be radiating near or above the Eddington limit.  This is consistent with the
possibility that Mrk 573 is a hidden narrow-line Seyfert 1 (NLSy1) galaxy (Ramos Almeida et al. 2008)\footnote{Unlike the
relatively broad polarized H$\alpha$ detected by Nagao et al. (2004), Ramos Almeida
et al. found a component of Pa$\beta$ with FWHM $\sim$ 1700 km$^{-1}$, which is consistent with limit of
$<$ 2000 km s$^{-1}$ for NLSy1s suggested by Osterbrock \& Pogge (1985).}. Interestingly,
Mrk 573 possesses a stellar bar in its nucleus, which appears to be common feature in the host galaxies of
NLSy1s (Crenshaw, Kraemer, \& Gabel 2003; Ohta et al. 2007).

\section{Photoionization Models}

\subsection{Model Input Parameters}

The  photoionization models used for this study were generated using the photoionization code
Cloudy, version 07.02.01 (last described by Ferland et al. 1998). We assumed an open, or ``slab'', geometry.
As per convention, the models are parameterized in
terms of $U$, the dimensionless ionization parameter, 
$U = Q/(4\pi r^{2}c n_{H})$
where $r$ is the radial distance of the absorber, $n_{H}$ is hydrogen number density, in units of cm$^{-3}$ and $Q = \int_{13.6 eV}^{\infty}(L_{\nu}/h\nu)~d\nu$,
or the number of ionizing photons s$^{-1}$ emitted by a source of luminosity
$L_{\nu}$, and $N_{H}$, the total hydrogen column density (in units of cm$^{-2}$).
  We assumed a spectral energy distribution (SED) similar to those in our previous
  studies (Kraemer \& Crenhsaw 2000a, 200b; Kraemer et al. 2000) in the form of a broken power law, such that $L_{\nu} \propto 
\nu^{\alpha}$ as follows: $\alpha = -0.5$ for energies $<$ 13.6 eV,
$\alpha = -1.5$ over the range 13.6 eV $\leq$ h$\nu$
$<$ 0.5 keV, and $\alpha = -0.8$ above 0.5  
keV. We included a low energy cut-off, of the power-law, at 1 eV and a high energy cutoff
at 100 keV. With this SED, scaled by  L$_{2-10 keV}$,   
$Q = 6 \times10^{54}$ photons s$^{-1}$.

Similar to our previous photo-ionization analyses (Kraemer \& Crenshaw 2000a,
2000b; Kraemer et al. 2000), we assumed
the following elemental abundances, in logarithm, relative to H 
by number: He: $-1.00$, C: $-3.47$, N: $-3.92$, O: $-3.17$, Ne: $-3.96$,
Na; $-5.69$, Mg: $-4.48$,  Al: $-5.53$, 
Si: $-4.51$,  P: $-6.43$, S: $-4.82$,  Ar: $-5.40$, 
Ca: $-5.64$,  Fe: $-4.40$, and Ni: $-5.75$. The heavy element abundances are roughly 1.4 times solar, except for
nitrogen which is twice solar (see Asplund, Grevesse, \& Sauval 2005), scaled in
the manner suggested by Groves, Dopita, \& Sutherland (2004a). 

Although some recent photoionization modeling studies have suggested that the dust/gas ratios in the NLR of Seyfert
galaxies are similar to that in the Galactic Interstellar Medium (ISM) (Groves, Dopita, \& Sutherland 2004b), 
these were not based on spectra obtained
with the 0\arcsecpoint05 spatial resolution of STIS. Groves et al. (2004c) analyzed STIS UV and optical spectra
of NGC 1068, and while finding that some line ratio comparisons, e.g C~IV $\lambda$1550/C~III] $\lambda$ 1909
versus C~IV $\lambda$1550/ He~II $\lambda1$1640, were consistent with the predictions of photoionization
models which included dust, they did not
attempt to model a full set of UV lines. By contrast, in our analysis of STIS spectra of NGC 4151 (Kraemer et al. 2000)
and NGC 1068 (Kraemer \& Crenshaw 2000a, 2000b), we determined that the dust/gas ratios in the inner $\sim$ 100 pc
were significantly less than that in the ISM. Two strong pieces of evidence for this conclusion were 1) several
regions with Ly$\alpha$/H$\beta$
$>$ 20 and 2) strong Mg~II $\lambda$2800 emission throughout the inner NLR. Ly$\alpha$, due to its large optical depth, is heavily suppressed
in dusty gas, while Mg~II $\lambda$2800 should be weak due to the depletion of magnesium onto dust grains. While the latter could
be ameliorated if the elemental abundances are super-solar, the relatively weak suppression of Ly$\alpha$ requires
fairly low dust/gas ratios. Based on these results, we included cosmic dust in the form of
graphite grains and silicate grains, with 50\% the dust/gas ratio of the Galactic Interstellar Medium (ISM). The 
inclusion of grains resulted in the following
depletions of elements from gas phase: C, 33\%; O 25\%; Mg, Si, Fe, Ca, Al, and Ni, 50\%.
While the wavelength-dependent extinction from such an anomalous grain population may differ greatly from
that in the ISM (see Crenshaw et al. 2001), the main effects of dust on our models are due to the depletion of elements
from gas phase and photo-electric heating by the grains.  

Prior to the detailed modeling, we can establish initial constraints on the model parameters. For example, the ratio of 
He~II~$\lambda$~4686/H$\beta$ is fairly high in the central knot (see Table 2), which
suggests the presence of matter-bounded gas (Kraemer \& Crenshaw 2000a, 2000b;  Kraemer et al. 2000). 
Also, since the [O~III] $\lambda$5007/$\lambda$4363 ratio is $>$ 50, it is likely that density of the
the gas in which these lines form is $<$ 10$^{5}$ cm$^{-3}$ (see Osterbrock \& Ferland 2006). 
  
Our approach in modeling the emission-line gas is to add as many individual components as necessary in order to
provide the best fit to the observed emission-line ratios. In previous studies, we have found that the line ratios
can be fit with one or two components. This approach differs from the so-called Locally Optimally-Emitting Cloud (LOC)
method suggested by Baldwin et al. (1995) and explored by Ferguson et al. (1997). Although one may argue that having a continuum
of physical conditions, as in LOC models, may be more realistic, we find that models with a small number of components
are sufficient in exploring the range in physical conditions within the NLR and can more clearly show the important effect of
absorption on the ionizing continuum (e.g. Collins et al. 2008). 

The presence of emission lines from such
a wide range in ionization and the relative blue-shifts detected in the high-ionization iron
lines (Mullaney \& Ward 2008; see also discussion in Section 2) are evidence for multiple distinct components of emission-line gas.
Further evidence for multiple components is the different spatial profiles of [O~II], [O~III], and [Fe~X] (see Figures 3 and 4), which
suggest the presence of at least 3 distinct emission components. 
Therefore, as we did in modeling the central emission-line knots in  
NGC 1068 (Kraemer \& Crenshaw 2000b) and NGC 4151 (Crenshaw \& Kraemer 2005), we assumed that the
emission-line gas is comprised of three components: 1) a moderately ionized component (MIDION), in which the bulk of the [O~III]
arises; 2) highly ionized gas (HIGHION), in which the
[Fe~X] forms; and 3) low-ionization gas (LOWION) in which the [O~II] and [N~II] form. 
Note that Binette, Wilson, \& Storchi-Bergmann (1996) performed a photoionization analysis of a subset of the emission-lines
detected in the ground-based spectra obtained
by Storchi-Bergmann et al. (1996) and found evidence for multiple emission components. However, the regions of the
NLR of Mrk 573
sampled by those spectra were much larger than those analyzed here.

Regarding the locations of these components, as shown in Figure 3, the central peaks of the [Fe~X] and [O~III] profiles
are coincident, therefore they are located at approximately the same distance from the AGN.  Hence, 
we have assumed that MEDION and HIGHION are  co-located, with different
densities corresponding to the difference in ionization state. 
Based on the strong [O~II] $\lambda$3727 emission and the
[S~II] 6731/6716 ratio, LOWION must have a density  n$_{H}$ $\leq$ 10$^{4}$ cm$^{-3}$
(Osterbrock \& Ferland 2006). The [O~II] $\lambda$3727/H$\beta$ ratio peaks at log$U \sim -4.0$  (Ferland \& Netzer
1983). For an unfiltered ionizing continuum, such a component would be
a distance of $\gtrsim$ 1 kpc from the AGN. Although the [O~II] profile is broader than [O~III], 
suggesting that there are contributions from emission-line gas over a range of radial distances, 
our extraction bin only extends $\sim$ 200 pc on either
side of the central knot. Therefore, the low-ionization gas would have to lie far out of the plane of the sky, which
seems unlikely given the biconical morphology of the NLR (see Figure 1). Instead, it is more probable that LOWION is
irradiated by a continuum which has been heavily filtered by intervening gas. Following Collins et al. (2008), in their modeling of the
Seyfert 2 galaxy Mrk 3, we have introduced such a filter characterized by log$U = -1.5$ and log($N_{H}/{\rm cm^{-2}}) = 21.6$ (see Figure 8).
We assumed that the filter lies close enough to the AGN to be obscured by dusty gas, hence is not detected in emission
in optical spectra.  We have also assumed that 
MIDION and HIGHION have the same density, which is consistent with the
density diagnostics discussed above.

\subsection{Model Results}

The final model parameters and the predicted emission-line ratios are listed in Table 3. 
We determined
$n_{H}$, $N_{H}$, and $U$ for MIDION based on the assumption that it is the source of
most of the  [O~III], [Ne~V] and He~II emission, and $U$ and $N_{H}$ for LOWION to provide
the best fit for the [O~II], [O~I], and [N~II] lines. The best fit for the low ionization lines
was obtained by allowing LOWION model to be radiation-bounded, unlike the other
two components.  Finally, as noted above, the parameters for HIGHION were 
optimized to produce the maximum [Fe~X]/H$\beta$ ratio. The composite model, using the
relative contributions to H$\beta$ listed in Table 3, provides a good fit to the observed
line ratios.

Regarding the dust/gas ratio, the [Fe~VII] lines are strongly
overpredicted compared to [Ne~V] in dust-free models and similarly underpredicted in models that used the
ISM dust/gas ratio. Hence, our assumption of an intermediate dust/gas ratio appears to be generally correct, although
the [Fe~VII] lines are still overpredicted, which suggests that the iron depletion may be
greater than 50\%. However, it should be noted that, if the iron depletion were changed to match the [Fe~VII] lines, it would
affect the match to [Fe~X]. Although it is possible that the dust/gas ratios are different in MEDION and HIGHION, given the
overall success of the models, we opted
against fine-tuning the model parameters to that extent. 

Based on the model parameters and estimate of the ionizing luminosity in Mrk 573, we derived
the radial distances from the AGN and covering factors for the three model components. The
predicted radial distance for MIDION and HIGHION is 55 pc, which is uncertain within a factor of $\sim$ 2 given the uncertainty in our estimate of the
ionizing luminosity (see Section 3). We scaled the transmitted
continuum from the filtering model to produce an ionization state for LOWION in which the low ionization lines dominate. This
resulted in a radial distance for LOWION of 168 pc.  We computed the emitting area for each component from the ratio of their contribution to the total H$\beta$ luminosity
divided by the predicted H$\beta$ flux from the ionized face of the slab (see
Table 3). The covering factors, which are the ratios of the emitting areas divided by the surface
areas of the spheres with radii centered at the AGN, 
and the depths (i.e., the radial extents) of the components, which are the ratios of of $n_{H}$ and $N_{H}$, are given in
Table 3. The small covering factors appear to be typical of the gas in the central regions of Seyfert 2s (Kraemer \& Crenshaw 2000a;
Collins et al. 2008). Since the depths are much less than the square root of the emitting areas, these
components would have the ``pancake'' structure predicted by Blumenthal \& Mathews (1979) for 
radiatively-driven clouds moving through an ambient medium, if they are
single clouds or filaments. However, another possibility is that each component
consists of a large number of roughly spherical clouds. 

\section{Obscuration of the Inner NLR}

It is instructive to compare our results for Mrk 573 with those from {\it HST} spectroscopic studies of the NLRs in other Seyfert galaxies.
We have analyzed STIS long-slit spectra of three other Seyferts: the Seyfert 2s NGC 1068 (Kraemer \& Crenshaw
2000a, 2000b) and Mrk 3 (Collins et al. 2005, 2008), and the Seyfert 1  NGC 4151 (Kraemer et al. 2000; Crenshaw \& Kraemer 2005, 2007).
We have also analyzed Faint Object Spectrograph (FOS) spectra of Seyfert 1 NGC 5548 (Kraemer et al. 1998). With the STIS spectra, we were able to resolve 
central emission line knots similar to that seen in Mrk 573. 
From our analysis of the central emission-line region in Mrk 573, we find conditions similar to the other two Seyfert 2. For
all three Seyfert 2s, the emission-line gas was found to be heterogeneous, with more
highly ionized components lying within the solid angle subtended by the collimated ionizing radiation, and the low-ionization 
components ionized by a heavily filtered continuum, which we argued places them outside the NLR bicone determined from
[O~III] images. This extended component of shielded gas is also present in NGC 4151 (Kraemer, Schmitt, \& Crenshaw 2008). However
there was no evidence for shielded gas in the central emission-line knot in NGC 4151, which is likely the result of the differences in inclination
between Seyfert 1s and 2s, as we discuss below.

The [O~III] $\lambda$5007/$\lambda$4363 ratio is sensitive to density as well as temperature, and at densities $\gtrsim$
10$^{6}$ cm$^{-3}$, the 5007 line is heavily suppressed due to collisional de-excitation of the $^{1}D_{2}$ level (Osterbrock \&
Ferland 2006).  The central
knots for the Seyferts 2 are consistent with low density gas, with 5007/4363 ratios of 52.4, in  Mrk 3; 47.8, in Mrk 573; and 35.2 for NGC 1068
(errors on the ratios are $\sim$ 20\%).
From the  models
we derived radial distances for the knots of 18 pc, 55 pc, and 35 pc, respectively. For NGC 4151, the 5007/4363
ratio is 13.8 and the emission-line gas is $<$ 10 pc from the AGN; in fact, there may be a strong
contribution from gas in the Intermediate Line Region (ILR), which is likely associated with the intrinsic UV and X-ray absorbers, at a radial distance of $\sim$ 0.1 pc (Crenshaw
\& Kraemer 2007). Although the central knot was not resolved in the FOS spectra of NGC 5548, the 5007/4363 was 10.5 and, therefore, 
we argued that much of the NLR emission arises in dense ($n_{H} \sim 10^{7}$ cm$^{-3}$) gas within 1 pc of the AGN (Kraemer et al. 1998), which, again likely
arises in an ILR
(Crenshaw et al. 2009). 

The radial distances we have derived are in general agreement with the unified model (Antonucci 1993). Specifically, our line-of-sight is
unobscured in the two Seyfert 1s, hence we are able to detect the innermost regions of the NLR. For the Seyfert
2s, the intervening obscuration prevents detection of gas close to the AGN. Our kinematic
modeling is also consistent with this picture, since we find that the inclination of the emission-line bicones
with respect to the plane of the sky is $\sim$ 5\deg~in Mrk 3 (Ruiz et al. 2001) and NGC 1068 (Das et al. 2006), and
$\sim$ 45\deg~ in NGC 4151 (Das et al.
2005).  
The global
covering factor of the obscuring material, i.e. the opening angle of the torus, has been
 estimated by comparing the relative number of Seyfert 1s and 2s (eg. Osterbrock \& Shaw 1988; Osterbrock
 \& Martel 1993). However, little
is known about the radial extent of the obscuring material.
In order to hide the inner emission-line regions in these three Seyfert 2s,
this material must extend tens of parsecs from the AGN.

\section{Summary} 

We have analyzed STIS long slit G430L and G750M spectra of the central [O~III] emission knot
in the Seyfert 2 galaxy Mrk 573. We have generated photoionization models of the
emission-line gas and have been able to match nearly all of the de-reddened emission line
ratios. We have found the following:

1. We compared the brightness profiles of the [Fe~X] $\lambda$6374, [O~III] $\lambda$5007,
and [O~II] $\lambda$3727 lines along the slit. The [Fe~X] line is just barely resolved, while the central
[O~III] emission is somewhat broader. The [O~II] profile is less centrally peaked and 
more extended than that of [O~III].  After extracting the central 1\arcsecpoint1~ of the spectra, we
found lines from a wide range in ionization state, suggesting a heterogeneous
emission-line region. Confirming the conclusions of Mullaley \& Ward (2007),
we found that the [Fe~X] line was broader than and blue-shifted compared to lower
ionization lines.
 
2. We estimated the 2-10 keV luminosity of Mrk 573 from the strength of the
[O~IV] 25.89 $\mu$m line, as described in Melendez et al. (2008a). Using the
bolometric correction from Awaki et al. (2001), Mrk 573 is radiating
at or above its Eddington Luminosity. This is consistent with the 
suggestion by Ramos Almeida et al. (2008) that Mrk 573 is a NLSy1, since
the accepted paradigm for NLSy1s posits that the AGN consists of
a relatively modest mass black hole, compared to broad-line Seyferts,
accreting matter at or above their Eddington limit (Pounds, Done, \& Osborne 1995).

3. We modeled the spectrum of the central knot using three components. MIDION and
HIGHION  are ionized by unfiltered radiation from the AGN. LOWION was irradiated 
by continuum radiation that has been heavily filtered by intervening gas,
which is close to AGN, hence likely obscured and not detected in emission. We assumed
that the emission-line gas contains cosmic dust, with a dust/gas ratio of 50\% that
of the Galactic ISM. The inclusion of dust in the models improved the overall fit to the
data, but the overprediction of the [Fe~VII] lines suggests that the dust/gas ratio, or
at least the depletion of iron from gas phase, may be somewhat larger.

4. The radial distance we derived for MIDION and HIGHION is $\sim$ 55 pc. The
fact that this region is tens of parsecs from the AGN is 
consistent with our modeling of the central emission-line regions in the
Seyfert 2 galaxies Mrk 3 and NGC 1068. On the other hand, we have derived
radial distances of a few parsces or less for the central knots
in the Seyfert 1 galaxies NGC 4151 and NGC 5548. Although the fact that we are
able to detect gas close the AGN is consistent with the predictions of the
unified model, the large radial distances of the central knots in these
Seyfert 2s requires that the obscuring material extends tens of parsecs from the
AGN.  We will address this issue in detail in a future paper.

\acknowledgments

This research made use of the NASA/IPAC Extragalactic Database (NED), which is
operated by the Jet Propulsion Laboratory, Caltech, under contract with NASA.
The observations used in this paper were obtained with the NASA/ESA Hubble
Space Telescope at the Space Telescope Science Institute, which is operated
by the Association of Universities for Research in Astronomy, Inc., under
NASA contract NAS5-26555. Basic research at the US Naval Research Laboratory (NRL)
is supported by the Office of Naval Research. Basic research in 
astronomy at NRL is supported by 6.1 base funding. We thank Gary Ferland for his
continued development and maintenance of CLOUDY. We thank the referee, Hagai
Netzer, for helpful and insightful comments.

\clearpage

\figcaption[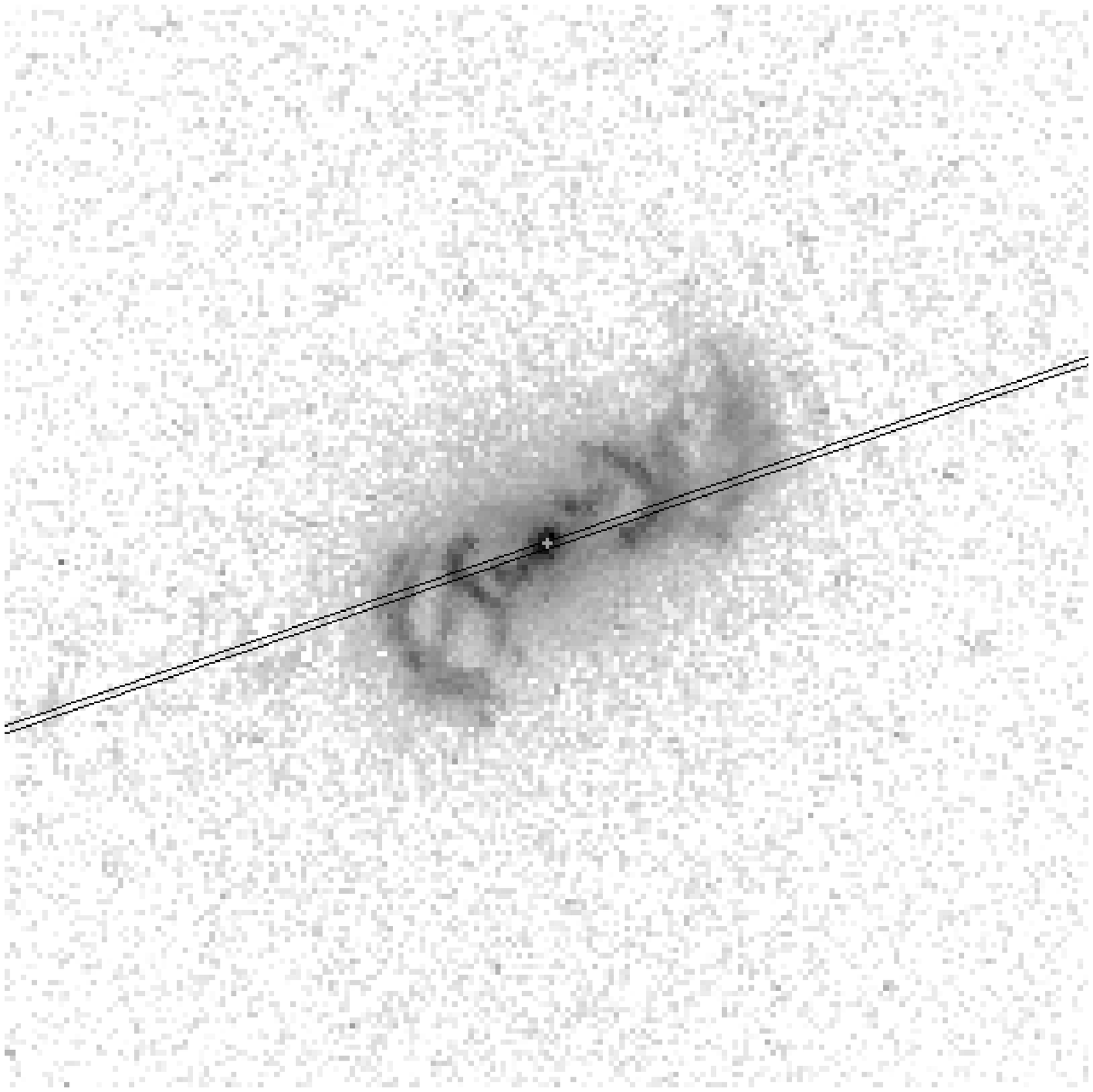]{WFPC2 [O~III] image of Mrk~573, showing the STIS slit
position. The image size is 20$''$ $\times$ 20$''$. North is up
and east is to the left. The peak of the continuum emission is identified
by a ``+''.}

\figcaption[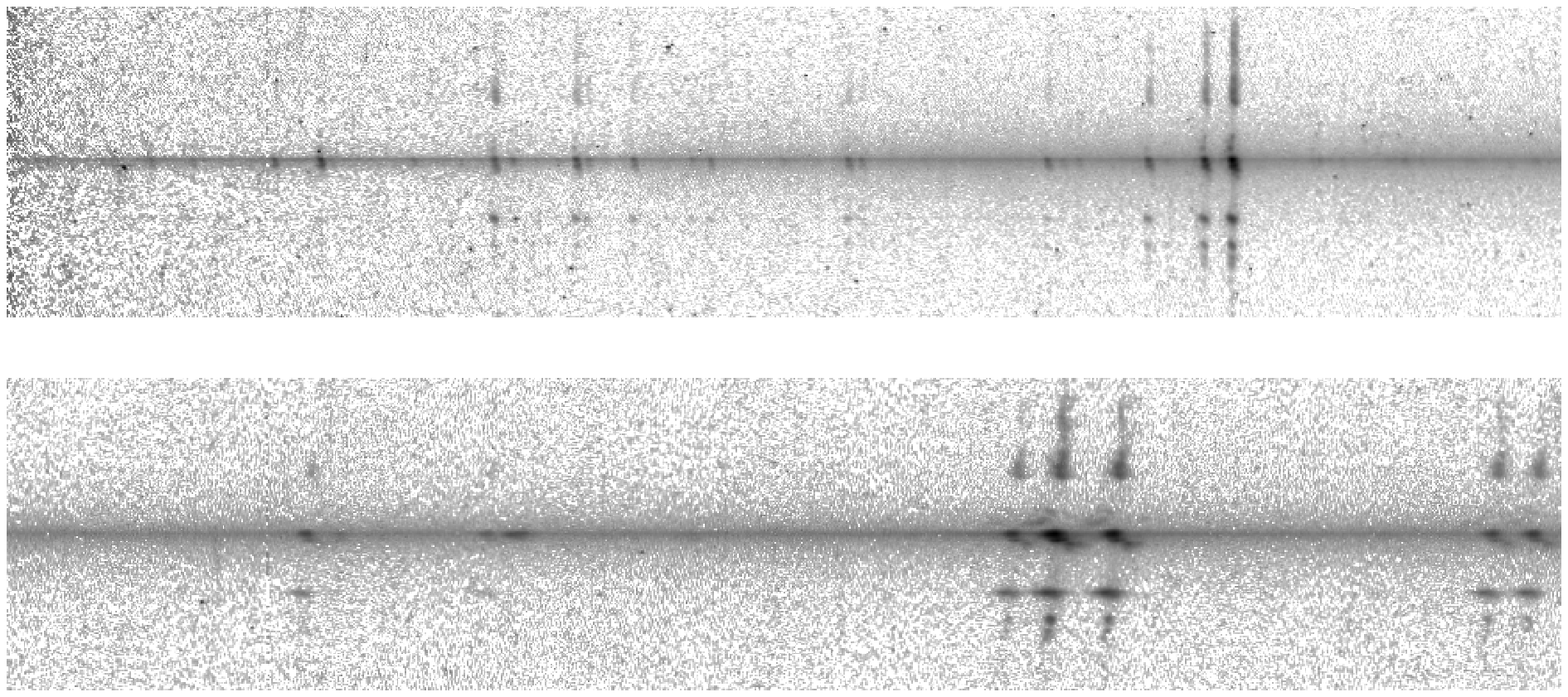]{Portions of the STIS G430L (top) and G750M (bottom)
spectral images. The images cover a span of 10$''$ in the spatial
direction. West is at the top and east is at the bottom in each image. The
dark horizontal lines are from the central continuum peak. The strongest
emission lines are [O~III] $\lambda$5007 in the top image and H$\alpha$ in
the bottom image.}

\figcaption[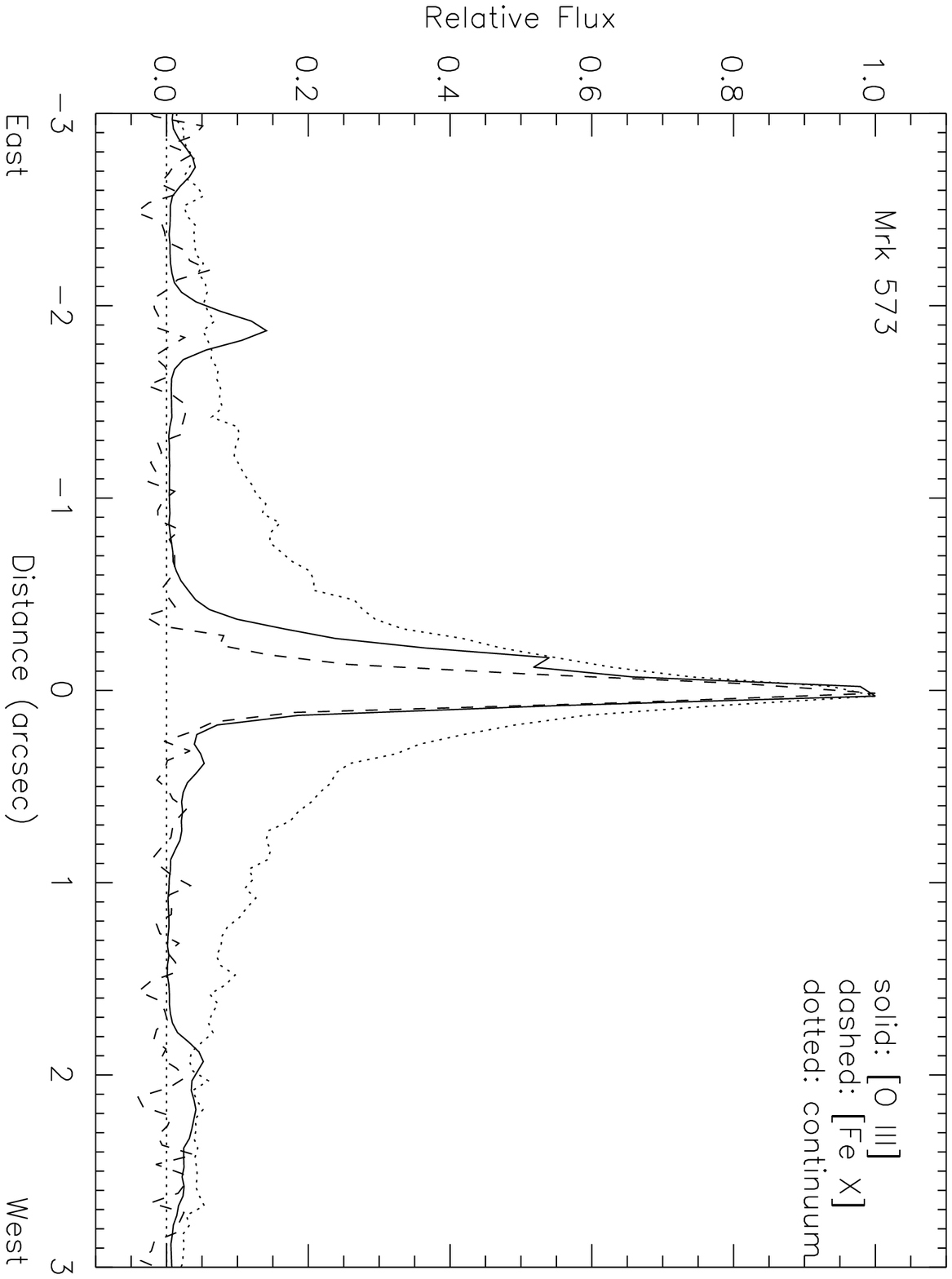]{Brightness profiles of [O~III] $\lambda$5007 and [Fe~X] $\lambda$6374 and the
continuum emission in the cross-dispersion direction. The profiles are
centered on the location of the continuum peak, and are normalized to
their peak fluxes.}

\figcaption[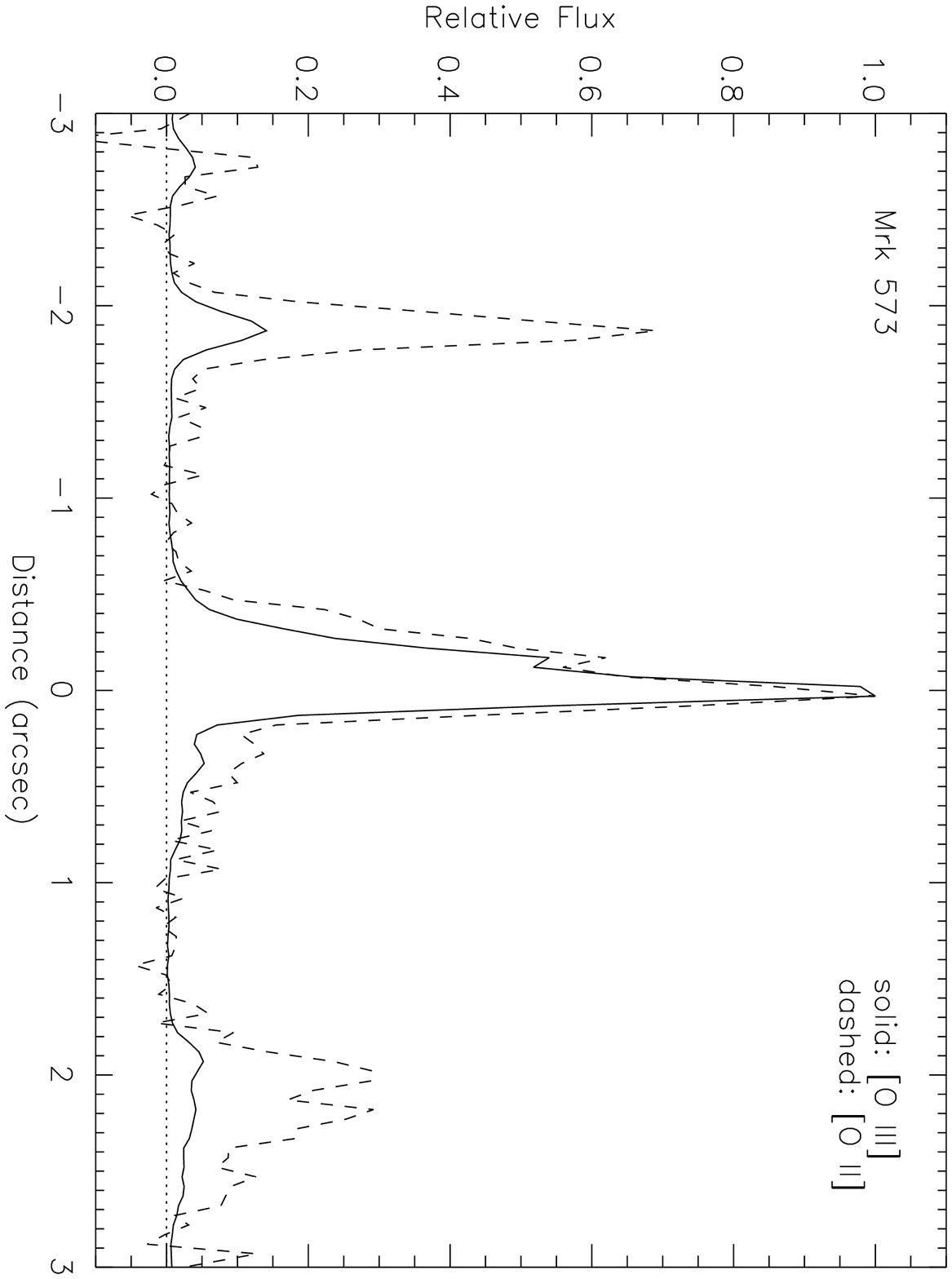]{Brightness profiles of [O~III] $\lambda$5007 and [O~II] $\lambda$3727 in the 
cross-dispersion direction. As in Figure 4, the profiles are
centered on the location of the continuum peak, and are normalized to
their peak fluxes.}

\figcaption[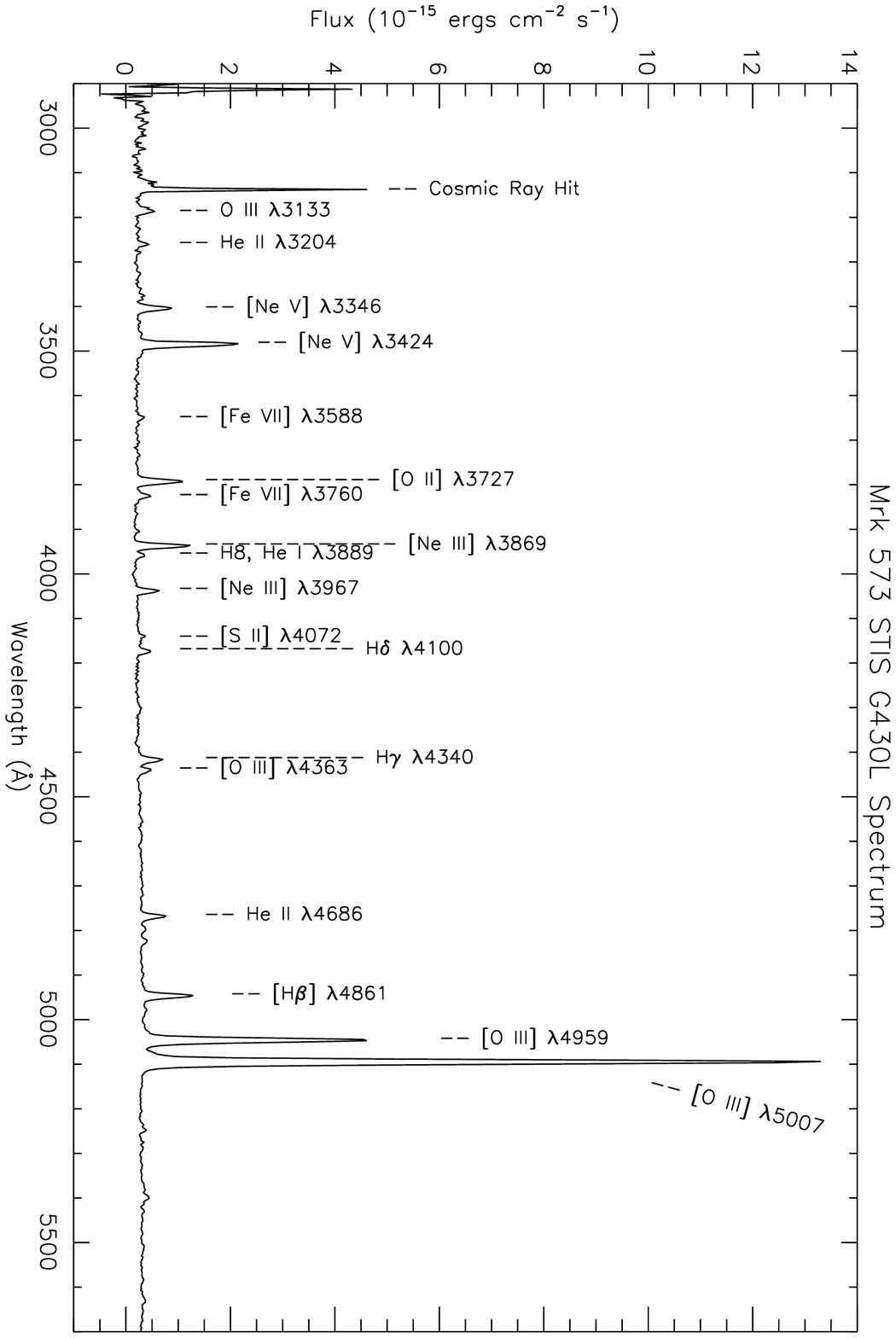]{STIS G430L spectrum of the central NLR knot in Mrk
573.}

\figcaption[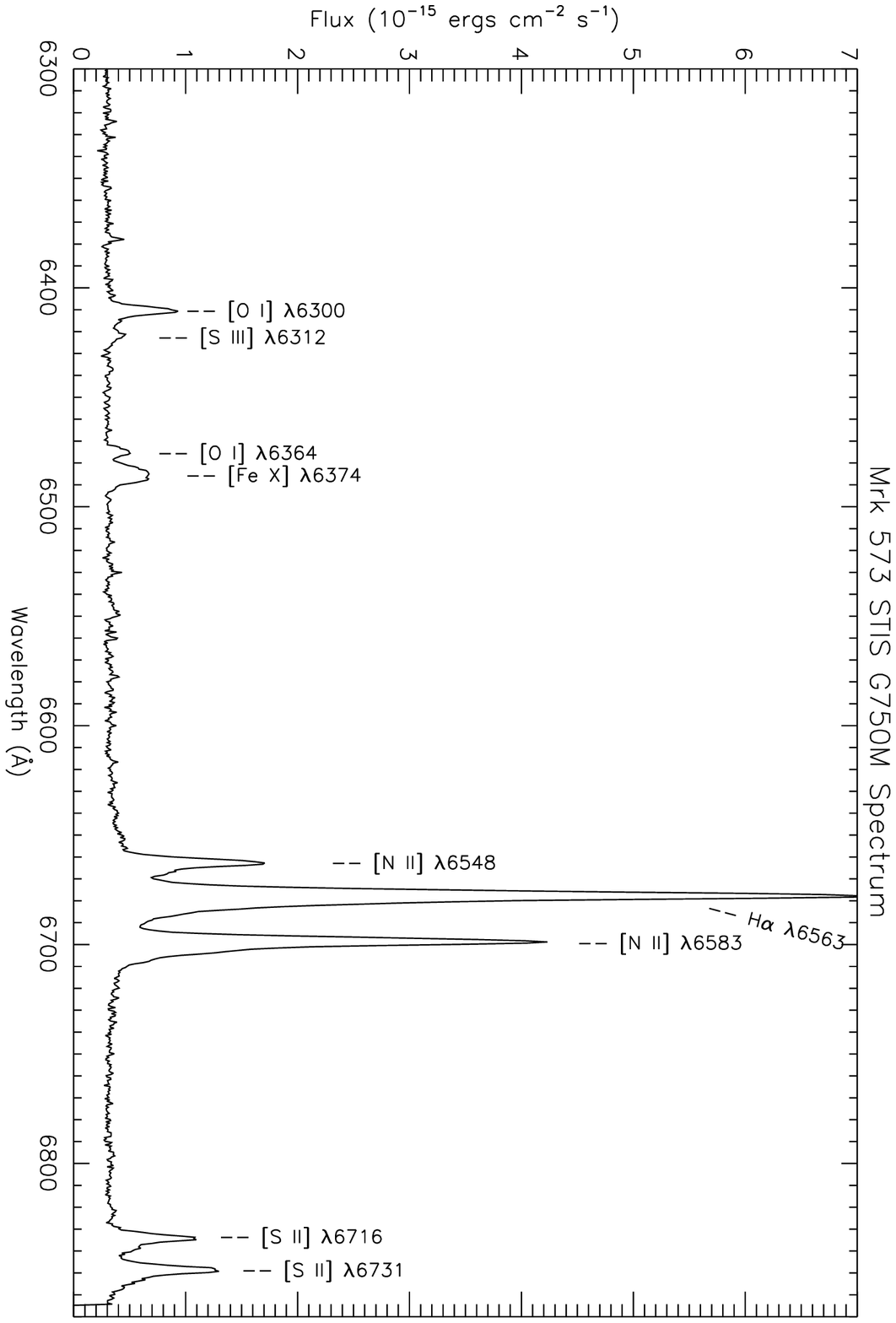]{STIS G750M spectrum of the central NLR knot in Mrk
573.}

\figcaption[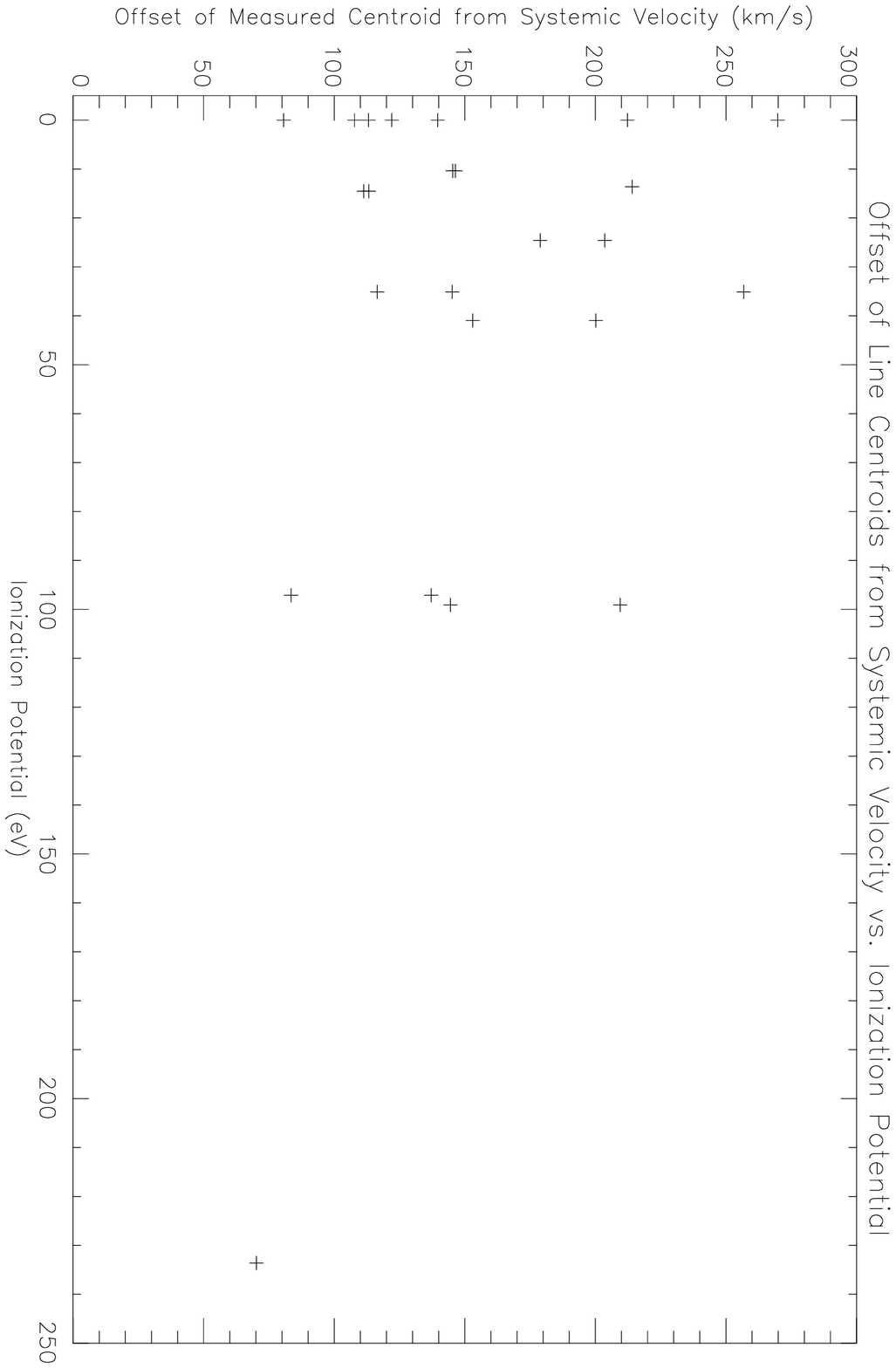]{Radial velocity centroid, relative to the systemic
redshift from H~I 21-cm emission, as a function of ionization potential for
the narrow emission lines.}

\figcaption[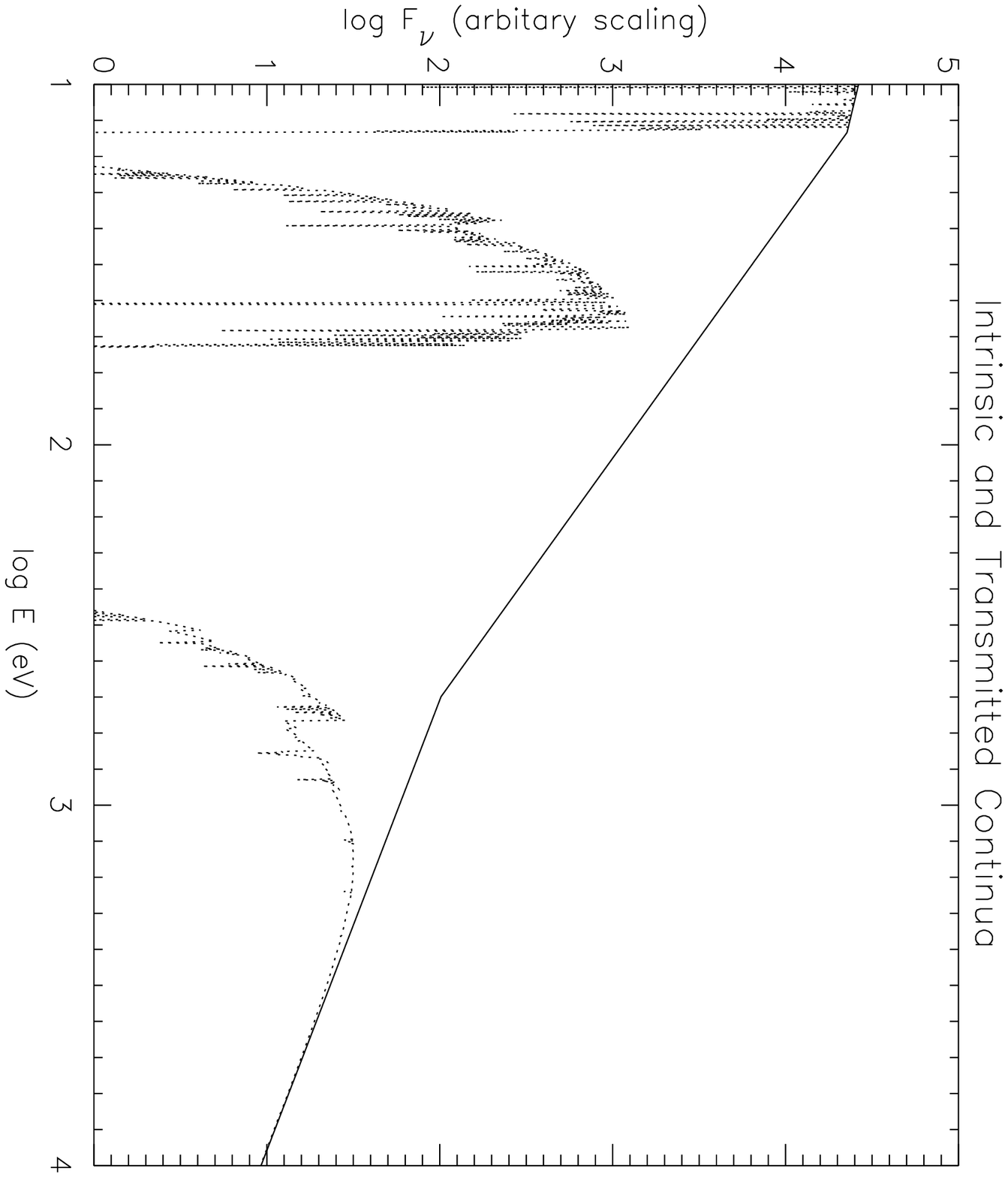]{Intrinsic (solid line) and transmitted continua (dotted line) for the 
filtering gas which lies between the AGN and LOWION. The model parameters for the filtering
gas are given in Section 4.1.}

\newpage
\begin{deluxetable}{ccrr}
\tablecolumns{6}
\footnotesize
\tablecaption{{\it HST}/STIS Observations of Mrk~573}
\tablewidth{0pt}
\tablehead{
\colhead{Grating} & \colhead{Coverage} & \colhead{Exposure} &
\colhead{Offset$^a$} \\
\colhead{} & \colhead{(\AA)} & \colhead{(sec)} & \colhead{(arcsec)}
}
\startdata
G750M &6300 -- 6970  &1080 &0.0 \\
G750M &6300 -- 6970  &1080 &0.25 \\
G750M &6300 -- 6970  &840 &$-$0.25 \\
G430L &2900 -- 5700  &840 &$-$0.25 \\
G430L &2900 -- 5700  &805 &0.0 \\
\enddata
\tablenotetext{a}{Offset position along the slit (a positive value
corresponds to upward displacement in the STIS spectral images).}
\end{deluxetable}

\begin{deluxetable}{lcccc}
\tabletypesize{\scriptsize}
\tablecaption{Mrk 573 Emission-Line Ratios Relative to H$\beta$ \label{tbl-2}}
\tablewidth{0pt}
\tablehead{
\colhead{} & \colhead{Observed Ratio} & \colhead{Dereddened} & \colhead{Velocity} & \colhead{ } \\
\colhead{Emission Line} & \colhead{to H$\beta$\tablenotemark{a}} & \colhead{Ratio\tablenotemark{b}} & \colhead{Offset\tablenotemark{c}} & \colhead{I.P.\tablenotemark{d}} 
}
\startdata
O III $\lambda$3133        & 0.27 $\pm$ 0.03 &  0.36 $\pm$ 0.04& --               &  35.12\\
He II $\lambda$3204        & 0.16 $\pm$ 0.03 &  0.21 $\pm$ 0.04& 79.7 $\pm$ 56.1  &  24.59\\         
$[$Ne V$]$ $\lambda$3346   & 0.65 $\pm$ 0.05 &  0.81 $\pm$ 0.07&  45.1 $\pm$ 19.4 &  97.12\\         
$[$Ne V$]$ $\lambda$3424   & 1.97 $\pm$ 0.08 &  2.43 $\pm$ 0.13&  4.0 $\pm$ 12.9  &  97.12\\         
$[$Fe VII$]$ $\lambda$3588 & 0.12 $\pm$ 0.02 &  0.14 $\pm$ 0.02& 211.9 $\pm$ 65.0 &  99.10\\         
$[$O II$]$ $\lambda$3727   & 0.95 $\pm$ 0.04 &  1.11 $\pm$ 0.06& 122.7 $\pm$ 0.8  &  13.62\\         
$[$Fe VII$]$ $\lambda$3760 & 0.30 $\pm$ 0.03 &  0.35 $\pm$ 0.04& 69.2 $\pm$ 30.6  &  99.10\\         
H9 $\lambda$3835           & 0.06 $\pm$ 0.01 &  0.07 $\pm$ 0.01&  --              &   0.00\\         
$[$Ne III$]$ $\lambda$3869 & 0.97 $\pm$ 0.04 &  1.11 $\pm$ 0.05& 62.1 $\pm$ 2.3   &  40.96\\         
H8, He I $\lambda$3889     & 0.19 $\pm$ 0.03 &  0.22 $\pm$ 0.03& 133.7 $\pm$ 27.3 &   0.00\\         
$[$Ne III$]$ $\lambda$3967 & 0.43 $\pm$ 0.04 &  0.49 $\pm$ 0.05& 111.3 $\pm$ 5.2  &  40.96\\         
$[$S II] $\lambda$4072     & 0.19 $\pm$ 0.02 &  0.21 $\pm$ 0.02&  --              &  10.36\\         
H$\delta$ $\lambda$4100    & 0.26 $\pm$ 0.03 &  0.29 $\pm$ 0.03& -20.5 $\pm$ 18.7 &   0.00\\         
H$\gamma$ $\lambda$4340    & 0.54 $\pm$ 0.04 &  0.58 $\pm$ 0.04& 91.7 $\pm$ 34.6  &   0.00\\         
$[$O III$]$ $\lambda$4363  & 0.26 $\pm$ 0.04 &  0.28 $\pm$ 0.04& 93.7 $\pm$ 17.6  &  35.12\\         
He II $\lambda$4686        & 0.48 $\pm$ 0.03 &  0.49 $\pm$ 0.03&  90.1 $\pm$ 5.0  &  24.59\\         
H$\beta$ $\lambda$4861     & 1.00 $\pm$ 0.05 &  1.00 $\pm$ 0.05&  31.6 $\pm$ 14.6 &   0.00\\         
$[$O III$]$ $\lambda$4959  & 4.44 $\pm$ 0.18 &  4.39 $\pm$ 0.18&  38.2 $\pm$ 7.7  &  35.12\\         
$[$O III$]$ $\lambda$5007  &13.63 $\pm$ 0.47 & 13.37 $\pm$ 0.46&  33.2 $\pm$ 5.9  &  35.12\\         
$[$Fe VII$]$ $\lambda$5159 & 0.07 $\pm$ 0.01 &  0.07 $\pm$ 0.01&  --              &  99.10\\         
$[$O I$]$ $\lambda$6300    & 0.29 $\pm$ 0.02 &  0.24 $\pm$ 0.02&  5.0 $\pm$ 10.3  &   0.00\\         
$[$S III$]$ $\lambda$6312  & 0.07 $\pm$ 0.01 &  0.06 $\pm$ 0.01&  --              &  23.34\\         
$[$O I$]$ $\lambda$6364    & 0.08 $\pm$ 0.01 &  0.07 $\pm$ 0.01&  --              &   0.00\\         
$[$Fe X$]$ $\lambda$6374   & 0.27 $\pm$ 0.01 &  0.23 $\pm$ 0.01& -18.4 $\pm$ 2.3  & 233.60\\        
$[$N II$]$ $\lambda$6548   & 0.70 $\pm$ 0.04 &  0.58 $\pm$ 0.04&  50.0 $\pm$ 1.8  &  14.53\\         
H$\alpha$ $\lambda$6563    & 3.53 $\pm$ 0.12 &  2.90 $\pm$ 0.14&  55.5 $\pm$ 2.2  &   0.00\\         
$[$N II$]$ $\lambda$6584   & 2.09 $\pm$ 0.13 &  1.71 $\pm$ 0.12&  99.7 $\pm$ 8.1  &  14.53\\
$[$S II$]$ $\lambda$6716   & 0.38 $\pm$ 0.02 &  0.31 $\pm$ 0.02&  63.1 $\pm$ 5.7  &  10.36\\
$[$S II$]$ $\lambda$6731   & 0.54 $\pm$ 0.02 &  0.44 $\pm$ 0.02&  54.3 $\pm$ 3.9  &  10.36\\
\enddata
\tablenotetext{a}{The integrated absolute flux of the H$\beta$ line in this spectrum is 11.9 $\pm$ 0.4 $\times10^{-15}$ ergs cm$^{-2}$ s$^{-1}$}
\tablenotetext{b}{Ratios dereddened using the standard Galactic reddening curve of Savage and Mathis 1979 and $E(B-V)=0.18 \pm 0.03$ mag, derived assuming the intrinsic ratio of H$\alpha$/H$\beta$=2.90.}
\tablenotetext{c}{Offset of line centroid from systemic velocity, in km s$^{-1}$. Those lines without measured offsets are either blended or too faint to measure the line's centroid.}
\tablenotetext{d}{Ionization potential, in eV.}
\label{tbl2}
\end{deluxetable}

\begin{deluxetable}{lccccc}
\tabletypesize{\scriptsize}
\tablecaption{Line Ratios from Individual Components and the Composite Model Compared with the Observed Ratios (Relative to H$\beta$)}
\tablewidth{0pt}
\tablehead{\colhead{Emission Line} 
&\colhead{MIDION$^{a}$} &\colhead{HIGHION$^{b}$} &\colhead{LOWION$^{c}$} &\colhead{Composite$^{d}$}
&\colhead{Observed$^{e}$}}
\startdata
~He~II $\lambda$3204 & 0.28 & 0.35 & $--$ & 0.22 & 0.21\\
~[Ne~V] $\lambda$3346 & 1.35 & 0.54 & $--$ & 0.93 & 0.81\\
~[Ne~V] $\lambda$3426 & 3.70 & 1.49 & $--$ & 2.55 & 2.43\\
~[Fe~VII] $\lambda$3588 & 0.48 & 0.51 & $--$ & 0.36 & 0.14\\
~[O~II] $\lambda$3727 & 0.03 & $--$ & 4.70 & 1.19 & 1.11\\
~[Fe~VII] $\lambda$3760 & 0.67 & 0.71 & $--$ & 0.51 & 0.35 \\
~H9 $\lambda$3835 & 0.08 & 0.06 & 0.07 & 0.08 & 0.07 \\
~[Ne~III] $\lambda$3869 & 1.59 & $--$ & 1.07 & 1.30 & 1.11\\
~H8, He~I $\lambda$3889 & 0.15 & 0.09 & 0.25 & 0.17 & 0.22\\
~[Ne~III] $\lambda$3967, H$\epsilon$ & 0.64 & 0.15 & 0.46 & 0.55 & 0.49\\
~[S~II] $\lambda$4072 & $--$ & $--$ & 0.71 & 0.18 & 0.22 \\
~H$\delta$ & 0.25 & 0.24 & 0.23 & 0.24 & 0.29\\
~H$\gamma$ & 0.46 & 0.47 & 0.44 & 0.46 & 0.58\\
~[O~III] $\lambda$4363 & 0.52 & $--$ & 0.01 & 0.34 & 0.28\\
~He~I $\lambda$4471 & 0.02 & $--$ & 0.05 & 0.03 & $--$ \\
~He~II $\lambda$4686 & 0.62 & 0.72 & $--$ & 0.48 & 0.49\\
~[O~III] $\lambda$4959 & 6.64 & $--$ & 0.50 & 4.44 & 4.39\\
~[O~III] $\lambda$5007 & 19.99 & $--$ & 1.49 & 13.37 & 13.37\\
~[Fe~VII] $\lambda$5159 & 0.31 & 0.25 & $--$ & 0.23 & 0.07\\
~[Fe~VI] $\lambda$5176 & 0.10 & $--$ & $--$ & 0.07 & $--$\\
~[N~I] $\lambda$5200 & $--$ & $--$ & 0.08 & 0.02 & $--$\\
~[O~I] $\lambda$6300 & $--$ & $--$ & 1.03 & 0.26 & 0.24\\
~[S~III] $\lambda$6312 & $--$ & $--$ & 0.04 & 0.01 & 0.06\\
~[O~I] $\lambda$6364 & $--$ & $--$ & 0.33 & 0.08 & 0.07\\
~[Fe~X] $\lambda$6374 & 0.01 & 2.26 & $--$ & 0.23 & 0.23\\
~[N~II] $\lambda$6548 & $--$ & $--$ & 1.85 & 0.46 & 0.58\\
~H$\alpha$ & 2.76 & 2.71 & 2.86 & 2.78 & 2.90\\
~[N~II] $\lambda$6584 & $--$ & $--$ & 5.45 & 1.36 & 1.71\\
~[S~II] $\lambda$6716 & $--$ & $--$ &  0.80 & 0.20 & 0.31\\
~[S~II] $\lambda$6731 & $--$ & $--$ & 1.38 & 0.35 & 0.44\\
\enddata
\tablenotetext{a} {~log$U = -1.25$; $n_{H}$ = 10$^{4.0}$ cm$^{-3}$; $N_{H} = 10^{21.35}$ cm$^{-2}$; $F_{H\beta} = 1.88$ ergs cm$^{-2}$ s$^{-1}$;
emitting area $=$ 10$^{39.43}$ cm$^{2}$; Depth $=$ 10$^{17.35}$ cm; Covering Factor $=$ 0.007.}
\tablenotetext{b} {~log$U = -0.47$; $n_{H} = 10^{3.2}$ cm$^{-3}$; $N_{H} = 10^{21}$ cm$^{-2}$; $F_{H\beta} = 0.08$ ergs cm$^{-2}$ s$^{-1}$;
emitting area $=$ 10$^{39.98}$ cm$^{2}$; Depth $=$ 10$^{17.8}$ cm; Covering Factor $=$ 0.026.}
\tablenotetext{c} {~log$U = -3.69$ (filtered); $n_{H} = 10^{4.0}$ cm$^{-3}$; $N_{H} = 10^{20.7}$ cm$^{-2}$ (radiation bounded); $F_{H\beta} = 0.03 $ ergs cm$^{-2}$ s$^{-1}$;
emitting area $=$ 10$^{40.8}$ cm; Depth $\geq$ 10$^{16.7}$ cm; Covering Factor $=$ 0.018.}
\tablenotetext{d} {Composite assuming following fractional contributions to H$\beta$: Comp 1, 0.65; Comp 2, 0.10; Comp 3, 0.25.}
\tablenotetext{e} {Dereddened: $E_{B-V} = 0.1794 \pm 0.031$.} 
\end{deluxetable}

\clearpage
\begin{figure}
\plotone{f1.eps}
\\Fig.~1.
\end{figure}

\clearpage
\begin{figure}
\plotone{f2.eps}
\\Fig.~2.
\end{figure}

\clearpage
\begin{figure}
\plotone{f3.eps}
\\Fig.~3
\end{figure}

\clearpage
\begin{figure}
\plotone{f4.eps}
\\Fig.~4
\end{figure}

\clearpage
\begin{figure}
\plotone{f5.eps}
\\Fig.~5
\end{figure}
\clearpage

\clearpage
\begin{figure}
\plotone{f6.eps}
\\Fig.~6
\end{figure}

\clearpage
\begin{figure}
\plotone{f7.eps}
\\Fig.~7
\end{figure}

\clearpage
\begin{figure}
\plotone{f8.ps}
\\Fig.~8
\end{figure}

\end{document}